\documentclass[aps,prl,twocolumn,nofootinbib,superscriptaddress]{revtex4-2}
\pdfoutput=1
\usepackage{graphicx}  
\usepackage{dcolumn}   
\usepackage{bm}        
\usepackage{amsmath,amssymb,amsxtra,bm,epsfig}   
\usepackage{float}
\usepackage[dvipsnames]{xcolor}
\usepackage{xspace}
\usepackage{hyperref}
\usepackage{orcidlink} 
\usepackage{cancel,soul,ulem}
\hypersetup{
	colorlinks=true,
	linkcolor=red,
	filecolor=magenta,
	urlcolor=ForestGreen,
	citecolor=RoyalBlue}

\usepackage[]{caption}
\newcommand{\bmt}{\begin{pmatrix}}
	\newcommand{\emt}{\end{pmatrix}}
\newcommand{\ba}{\begin{array}{c}}
	\newcommand{\ea}{\end{array}}
\newcommand{\be}{\begin{equation}}
	\newcommand{\ee}{\end{equation}}
\newcommand{\bea}{\begin{eqnarray}}
	\newcommand{\eea}{\end{eqnarray}}

\newcommand{\bi}{\begin{itemize}}
	\newcommand{\ei}{\end{itemize}}
\newcommand{\baz}{\begin{array}{cc}}
\newcommand{\besub}{\begin{subequations}}
\newcommand{\eesub}{\end{subequations}}

\begin{document}
	\title{Delayed Charged Lepton Yukawa Equilibration in Minimal Seesaw}

	\author{Rishav Roshan\,\orcidlink{0000-0002-8545-4188}}
	\email{r.roshan@soton.ac.uk}
	\affiliation{School of Physics and Astronomy, University of Southampton,\\ Southampton, SO17 1BJ, U.K.}
    
	\author{Sudipta Show\,\orcidlink{0000-0003-0436-6483}}
	\email{sudiptas@iitk.ac.in}
	\affiliation{Department of Physics, Indian Institute of Technology Kanpur, Kanpur 208016, India}

		\begin{abstract}

           While most studies of the \textit{minimal type-I seesaw} neglect the heaviest decoupled right-handed neutrino, assuming negligible contributions to neutrino masses and leptogenesis, we demonstrate that its cosmological role can be significant. When long-lived, the presence of this particle can substantially modify the charged lepton Yukawa equilibration temperature in the early universe, necessitating reassessment of lepton flavor effects in thermal leptogenesis and potentially shifting flavor regime boundaries. Additionally, we identify experimental probes, including neutrino oscillation measurements and gravitational wave observations to investigate this scenario. More broadly, the resulting modification of the cosmic expansion history carries implications beyond leptogenesis, impacting dark matter production and evolution, the matter-antimatter asymmetry generation, and the primordial gravitational wave spectrum. 
\end{abstract}

		\maketitle
        
The type-I seesaw mechanism~\cite{Minkowski:1977sc,Yanagida:1979as,Yanagida:1979gs, GellMann:1980vs,Mohapatra:1979ia,Schechter:1980gr,Schechter:1981cv,Datta:2021elq} stands as one of the most compelling and theoretically well-motivated approaches to understanding the origin of neutrino masses, a problem that continues to pose a fundamental challenge to the Standard Model (SM) of particle physics. In this realization, the SM field content is extended by introducing three heavy right-handed neutrinos (RHNs) $N_i$, which are singlets under the SM gauge group, and that couple to the 
SM lepton doublet ($\ell_L^T = (\nu_L, e_L)$) and the Higgs doublet ($H$). The corresponding Lagrangian is then given by
\begin{align}
	-\mathcal{L}= \overline{\ell_L}_{\alpha} (Y_{\nu})_{\alpha i} \tilde{H} N_{i}+ \frac{M_{i}}{2}  \overline{N_{i}^c} N_i+ h.c.,
\end{align}

\noindent in the charged lepton diagonal basis with $\alpha = e, \mu, \tau$ and $i = 1, 2, 3$. Interestingly, these RHNs have strong theoretical motivation beyond phenomenology. They naturally arise in grand unified theories (GUTs) such as $SO(10)$ or $E_6$ and are crucial for anomaly cancellation in extensions of the SM that include additional gauge symmetries like $U(1)_{\rm B-L}$ or $SU(2)_R$. 

It is well known that the observed neutrino oscillation data can be successfully  accommodated with just two RHNs, representing a limiting case of the three RHN framework in which  one RHN effectively decouples from the seesaw mechanism via an extremely large mass, vanishingly small Yukawa couplings, or both. This minimal setup inherently predicts one massless active neutrino and has been extensively studied in the literature~\cite{King:1999mb, Antusch:2011nz, Li:2017zmk,King:2025eqv}. However, the cosmological role of the decoupled RHN, say $N_3$, has largely been overlooked. 

In this letter, we investigate precisely this question. We demonstrate that the decoupled RHN can drive an epoch of early matter domination (MD), significantly modifying the expansion history of the Universe. This modified cosmological background has far-reaching consequences for a wide range of SM interactions, most notably the equilibration dynamics of charged lepton Yukawa interactions, which constitutes the primary focus of this work. Beyond this, the early MD epoch can also leave observable imprints on other cosmological phenomena, including the generation of the matter-antimatter asymmetry, the production and evolution of dark matter, and the generation and propagation of gravitational waves (GWs). Thus, although our analysis centers on the impact of delayed charged lepton Yukawa equilibration on leptogenesis, the underlying mechanism is far more general, bearing relevance for a broad class of particle physics and cosmological scenarios.

As a concrete example, we first discuss how the decoupled RHN modifies the cosmological history, thereby delaying the equilibration of charged lepton Yukawa interactions, and subsequently examine its impact on leptogenesis\footnote{Beyond the neutrino mass problem, the presence of two heavy RHNs inherent to this framework also provides a natural setting for leptogenesis~\cite{Antusch:2011nz}.}, driven by the out-of-equilibrium decay of the lightest RHN, $N_1$~\cite{Buchmuller:2005eh, Davidson:2008bu}. Assuming all the three RHNs to be in equilibrium \footnote{This can be naturally realized, for example, in a gauged $B-L$ setup, where the associated gauge interactions rapidly bring the RHNs into equilibrium with the thermal plasma. Furthermore, the long lifetime needed for the $N_3$ dominated cosmological epoch would remain unspoiled by the $B-L$ gauge interactions. } when produced after the Universe was reheated, $N_3$ later decouples due to its extremely weak (but non-zero) interaction with the SM lepton and Higgs doublet. These suppressed couplings simultaneously ensure $N_3$ is long-lived, allowing it to dominate the cosmic energy density and drive an epoch of MD, while also yielding a tiny but non-zero mass for the lightest active neutrino (denoted by $m_\ell$). To achieve this, we consider the following structure of the neutrino Yukawa matrix\footnote{A similar structure of Yukawa was also considered in \cite{Datta:2021elq} but in the context of dark matter.} at the leading order
\begin{align}
		Y_{\nu}=\left(
		\begin{array}{ccc}
		y_{e1} & y_{e2} & \epsilon_1 \\
		y_{\mu 1} & y_{\mu 2} & \epsilon_2 \\
		y_{\tau 1} & y_{\tau 2} & \epsilon_3 \\
		\end{array}
		\right), \label{eq:yukawa}
	\end{align}
where $\epsilon_{i=1,2,3} \ll 1$. While the nearly vanishing right column of the Yukawa matrix ensures that $N_3$ remains effectively decoupled, the $N_{1,2}$ along with the first two columns of $Y_{\nu}$ generate light neutrino mass to explain the neutrino oscillation data via the seesaw mechanism. The entries of $Y_{\nu}$ can be written using the Casas-Ibarra (CI) parametrization \cite{Casas:2001sr}: 
\begin{align}
		Y_\nu= \frac{\sqrt{2}}{v} U D_{\sqrt{m}} R^T D_{\sqrt{M}},
		\label{CI para}
	\end{align}
where $v=246$ GeV, $U$ is the PMNS \cite{Zyla:2020zbs} mixing matrix diagonalizing $m_{\nu} =  Y_{\nu} M^{-1}_i Y^T_{\nu}v^2/2 $, $D_{m} ~(D_{M})$ is the diagonal active neutrino (RHN) mass matrix and $R$ is a complex orthogonal matrix chosen to be of the form, 
\begin{align}
		R=\left(
		\begin{array}{ccc}
		0 & \cos \theta_R & \sin \theta_R \\
		0 & -\sin \theta_R & \cos \theta_R \\
        1 & 0 & 0 \\
		\end{array}
		\right), \label{eq:R}
	\end{align}

\noindent where $\theta_R=a+ib$ is a complex angle in general. We employ the best fitted values \cite{Fogli:2006fw} of mixing angles, CP phase, as well as mass-square differences to define the $U$ and $D_{\sqrt{m}}$ assuming a normal hierarchy for the active neutrino masses. The smallness of $\epsilon_i$ can arise from three sources: a tiny $m_\ell$ (lightest active neutrino mass), an additional rotation angle $\varphi$ applied to $R$, or both effects combined. For concreteness and without loss of generality, we focus on the small $m_\ell$ scenario in our analysis, noting that equivalent results are obtained in the $\varphi$ rotation approach. Following Eq.(\ref{CI para}), this choice yields $\epsilon_i \propto \sqrt{m_\ell M_3}$ \footnote{Note that, this relation also hold for the inverted hierarchy with an appropriate choice of $R$.}. We emphasize that $N_3$ phenomenology is largely insensitive to the left two columns of $Y_{\nu}$.

We now proceed to investigate the cosmological evolution leading to early MD driven by the long-lived heavy neutrino $N_3$ under the assumption that this species was in thermal equilibrium at the time of its production. The relevant dynamics are governed by the following coupled Boltzmann equations (BEs):

\begin{align}
a\frac{d\rho_{N_3}}{da}+F\rho_{N_3}=-\frac{\Gamma_{N_3}}{\mathcal{H}}\rho_{N_3} \nonumber\\
a\frac{d\rho_R}{da}+4\rho_R=+\frac{\Gamma_{N_3}}{\mathcal{H}}\rho_{N_3}
\label{BE}
\end{align}

\noindent where $\rho_R$ represents the energy density of SM radiation and $\rho_{N_3}$ and $\Gamma_{N_3}= \frac{(Y_\nu^\dagger Y_\nu)_{33}}{8\pi}M_3$ refer to the the energy density and decay width of $N_3$ respectively. $\mathcal{H}=\sqrt{\rho_{N_3}+\rho_R}/3M_P^2$ denotes the expansion rate of the universe with $M_P(=2.4\times10^{18}~\text{GeV})$ being the reduced Planck mass. The factor $F$ dictates the behaviour of $\rho_{N_3}$'s evolution where $F=4~(3)$ when $N_3$ is relativistic (non-relativistic).

\begin{figure}[h]
	\includegraphics[width=0.95\linewidth]{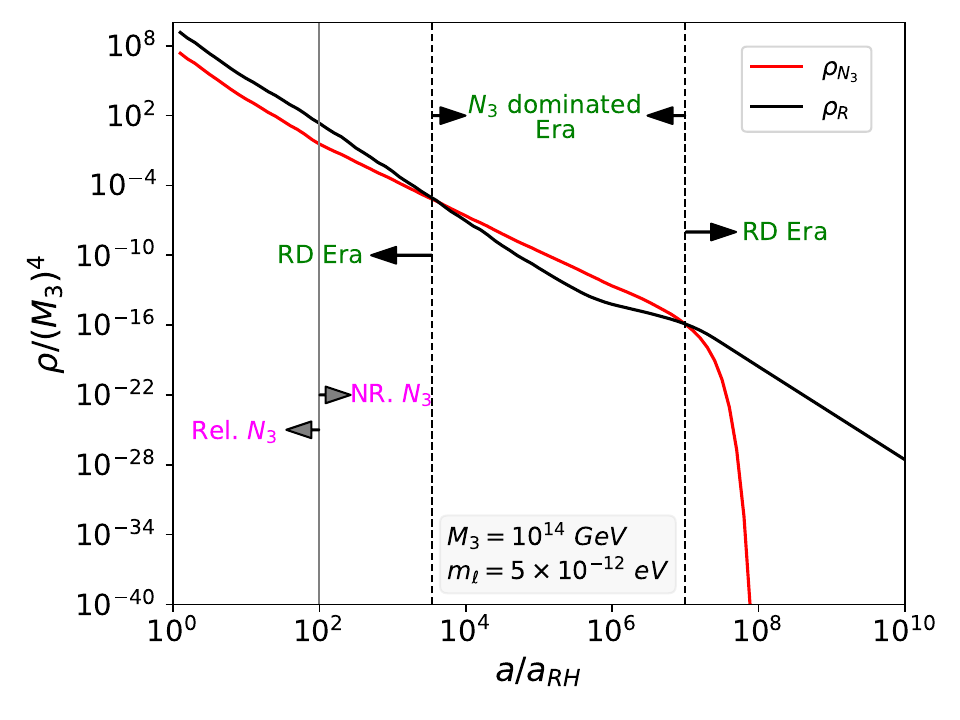}
	\caption{ Evolution of the various energy densities as a function of the relative scale factor, $a/a_{\text{RH}}$.}\label{fig:energy_density}
\end{figure}

For illustration, in Figure~\ref{fig:energy_density}, we present a numerical solution to the above BEs showing how $N_3$ induces an era of early matter domination. The plot is obtained using $m_\ell = 5\times 10^{-12}$  eV and $M_3 = 10^{14}$ GeV. The red (black) line shows the evolution of the $N_3$ (radiation) energy density with the scale factor $a$ (normalized with respect to $a_{\text{RH}}$ defined at reheating temperature ($T_{\rm RH}=10^{16}$ GeV), assuming an instantaneous reheating after the end of inflation as an initial condition). Initially, $N_3$ behaves as radiation with energy density scaling as $\rho_{N_3} \propto a^{-4}$. Once the universe temperature ($T$) drops to $T \sim M_3$, the $N_3$ becomes non-relativistic and its energy density starts following matter-like scaling, $\rho_{N_3} \propto a^{-3}$. Consequently, the universe enters an early matter-dominated era that persists until $N_3$ decays completely.

One immediate consequence of this early matter dominated era is a shift in the equilibration temperature (ET) $T^*$ of the SM charged Yukawa interactions \footnote{A similar analysis was done in \cite{Datta:2022jic, Datta:2023pav} but in the context of extended reheating.}. The equilibration of these interactions plays a non-trivial role when studying the effect of different lepton flavors in leptogenesis \cite{Barbieri:1999ma, Nardi:2005hs, Abada:2006fw, Nardi:2006fx, Blanchet:2006be, Dev:2017trv, Datta:2021gyi}. Specifically, when a given charged lepton Yukawa interaction is in equilibrium at temperature in a radiation-dominated (RD) era denoted by  $T^*_{0(\alpha)}$ , the lepton doublets ${\ell}_{L\alpha}$ from RHN decays can interact with the corresponding $e_{R_\alpha}$ states in the bath. These interactions destroy the quantum coherence between different lepton doublet states, which in turn dictates the number of orthogonal flavor alignments in that particular temperature regime. 
\begin{figure}[h]
	\includegraphics[width=0.95\linewidth]{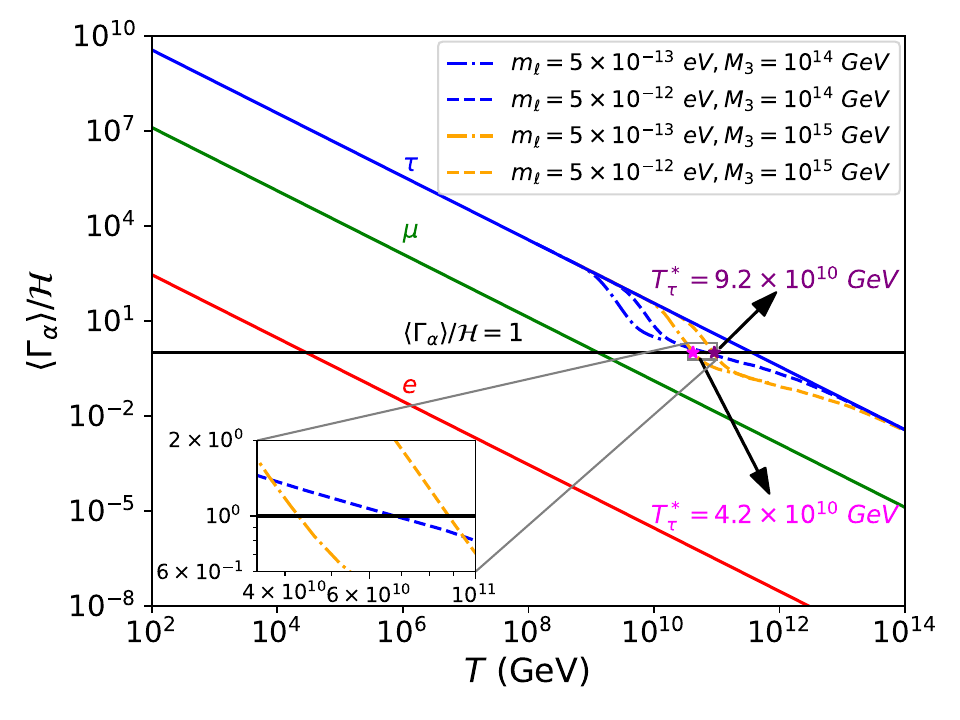}
	\caption{ Variation of $\langle \Gamma_\alpha \rangle/\mathcal{H}$ vs. $T$ for standard (solid
lines) and modified (dot-dashed and dashed) scenarios.}\label{gammabyH}
\end{figure}

In the standard RD era, the ET at which a specific $e_{R_\alpha}$ equilibrates with corresponding lepton doublet can be naively estimated by comparing the Higgs boson decay (and inverse decay) rate $\langle \Gamma_{\alpha} \rangle$ with that of the Hubble expansion $\mathcal{H}$~\cite{Cline:1993vv, Cline:1993bd}. In the case of $\tau_R$ in the RD universe, this equilibration condition reads
\begin{equation}
	\frac{\pi Y^2_{\tau}}{192 \zeta(3)} \frac{m^2_h (T)}{T} = 0.066 {g_*}^{1/2}\frac{T^2}{M_P},
	\label{eq-temp}
\end{equation}
where $\mathcal{H}(T) = 0.066 {g_*}^{1/2}\frac{T^2}{M_P}$ is used, $g_*$ represents effective number of relativistic degrees of freedom. The thermal Higgs mass is given by $m_h (T)\simeq0.6T$~\cite{Weldon:1982bn, Quiros:1999jp, Senaha:2020mop}.  Solving Eq.~(\ref{eq-temp}) yields $T^*_{0({\tau})} = 5 \times 10^{11}$ GeV, corresponding to the intersection point between the solid blue curve (standard cosmology) and the horizontal black line at $\langle \Gamma{\tau} \rangle/{\mathcal{H}} = 1$ in Figure~\ref{gammabyH}. The analogous equilibration temperatures for $\mu_R$ and $e_R$ are $T^*_{0({\mu})} = 10^9$ GeV and $T^*_{0(e)} = 5 \times 10^4$ GeV, respectively.  

As discussed, the standard evaluation of ET assumes RD throughout. In this letter, we demonstrate that a decoupled RHN in the seesaw mechanism can significantly alter the expansion history of our early universe and shift the ET of the charged Yukawa interactions by inducing an early MD. We denote the shifted ET for different lepton flavor as $T^*_{\alpha}$ just to distinguish it from the ET obtained in the standard RD era ($T^*_{0(\alpha)}$) and illustrate this in Figure~\ref{gammabyH}. The modified $\langle \Gamma{\tau} \rangle/{\mathcal{H}}$ shown by the dashed and dot-dashed lines in the plot is obtained by appropriately writing $\mathcal{H}$ in terms of $\rho_R$ and $\rho_{N_3}$ (in contrast to the standard scenario, which only considers $\rho_R$) and substituting it in the set of BEs given in Eq. \eqref{BE}. The solution is then plugged into the relation.
\begin{equation}
	T(a) = \left[ \frac{30}{g_*(a)\pi^2}\right]^{1/4} \rho^{1/4}_R(a)
    \label{temp_eq}
\end{equation}
connecting the temperature $T(a)$ to $\rho_R$. The presence of $\rho_{N_3}$ in $\mathcal{H}$ induces a modified expansion of the universe simultaneously affecting Eq. \eqref{eq-temp} following Eq.~\eqref{temp_eq}. 

In Figure~\ref{gammabyH}, for $M_3=10^{14}$ GeV and $m_\ell= 5\times10^{-12}$ eV (dashed blue line), we observe a departure from the solid blue line. The departure begins at a high temperature when the universe enters the $N_3$-driven matter-dominated phase and ends at a lower temperature when $N_3$ completely decays, after which the RD era resumes. Consequently, $\langle\Gamma{\tau} \rangle/{\mathcal{H}}=1$ is achieved at $T^*_{{\tau}} = 7 \times 10^{10}$ GeV marking a clear shift from the standard RD scenario where $T^*_{0({\tau})} = 5 \times 10^{11}$ GeV. Reducing $m_\ell$ to $5\times 10^{-13}$ eV (dot-dashed blue line) while keeping $M_3$ fixed does not change the initial point where $\langle\Gamma_{\tau} \rangle/{\mathcal{H}}$ departs from the standard scenario, but delays the merger with the solid blue line at lower temperature. Such a delay occurs because a smaller $m_\ell$ reduces the $N_3$ decay width, extending the period of the matter-dominated era. As a result, radiation domination after $N_3$ decay begins at a later epoch (lower temperature).

On the other hand, increasing the value of $M_3$ to $10^{15}$ GeV (orange lines) while keeping the same two values of $m_\ell$ reduces the life-time of $N_3$, shifting the variation of $\langle\Gamma_{\tau} \rangle/{\mathcal{H}}$ to higher temperatures. In this case, the cross-over $\langle\Gamma{\tau} \rangle/{\mathcal{H}}=1$ happens at two different values of temperature: $T^*_{{\tau}} \simeq 9.2 \times 10^{10}$ GeV for $m_\ell= 5\times10^{-12}$ eV (dashed orange line) and $T^*_{{\tau}} = 4.2 \times 10^{10}$ GeV for $m_\ell\simeq 5\times10^{-13}$ eV (dot-dashed orange line). The latter represents an order-of-magnitude shift compared to the standard scenario. Similar shifts in $T^*_{{e}}$ and $T^*_{{\mu}}$ can also be achieved by appropriately choosing $M_3$ and $m_\ell$. 

The modified ET has immediate implications for leptogenesis. As discussed, if the universe is dominated by radiation as in the vanilla leptogenesis, the ET for $\tau_R,\mu_R$ and $e_R$ flavors are $T^*_{0({\tau})} = 5 \times 10^{11}$ GeV, $T^*_{0({\mu})} = 10^9$ GeV and $T^*_{0(e)} = 5 \times 10^4$ GeV, respectively. In this case, if the lightest RHN $N_1$ with mass $M_1=10^{11}$ GeV decays to produce lepton asymmetry, one needs to properly take into account the effect of the equilibration of $\tau_R$ while solving the BEs and studying the evolution of the lepton asymmetry with the expansion of the universe. On the contrary, if the $N_1$ is heavier, $i.e~ M_1>5\times 10^{11}$ GeV, the quantum coherence between different lepton doublet states is not broken and one can stick to the unflavored leptogenesis scenario \cite{Fukugita:1986hr, Luty:1992un, Pilaftsis:1997jf, Konar:2020vuu}.

Unlike the standard leptogenesis proceeding in the RD era, an early MD driven by $N_3$ modifies the ETs as shown in Figure~\ref{gammabyH}. With $M_1 = 10^{11}$ GeV and with appropriate choices of $M_3$ and $m_\ell$, now the system remains in the unflavored leptogenesis regime due to the shift of $T^*_{0({\tau})}\to T^*_{{\tau}}$. Therefore, we proceed to evaluate the $B-L$ asymmetry using the following Boltzmann equation \cite{Buchmuller:2004nz}, 
\begin{align}
a\frac{d \rho_{N_1}}{da}+F\rho_{N_1}= -\frac{\langle\Gamma_{N_1}\rangle }{\mathcal{H}}(\rho_{N_1}-\rho_{N_1}^{\text{eq}}),\nonumber
\end{align}
\begin{align}
	a\frac{d n_{\Delta}}{da}+3n_\Delta=-\frac{\langle\Gamma_{N_1}\rangle}{\mathcal{H}}\left[\frac{\varepsilon_1}{M_1}(\rho_{N_1}-\rho_{N_1}^{\rm{eq}})+\frac{n_{N_1}^{\rm{eq}}}{2 n_\ell^{\rm{eq}}}n_{\Delta}\right],
    \label{BE_asy}
\end{align}
\noindent with $\rho_{N_1}$ ( $\rho_{N_1}^{\rm eq}$) denoting the energy density (equilibrium energy density) of $N_1$, $\langle \Gamma_{N_1}\rangle$ denoting the thermally averaged decay width of $N_1$, $n_{\Delta} = n_{B-L}$ and the CP asymmetry produced from the decay of the lightest RHN $N_1$ is given by \cite{Covi:1996wh} :
\begin{equation}\label{eq:cpasym}
	\varepsilon_1=\frac{1}{8\pi} \frac{\text{Im}[({Y}_\nu^\dagger {Y}_\nu)^2_{12}]}{({Y}_\nu^\dagger {Y}_\nu)_{11}}{\cal F}\left(\frac{M_2^2}{M_1^2}\right)\,,
\end{equation}
where ${\cal F}(x)\simeq 3/2\sqrt{x}$ for hierarchical RHNs. Note that, in contrast to $N_3$, the RHNs $N_{1,2}$ never dominate the energy budget due to their faster decay rates and therefore do not affect ETs. We would also like to point out that, the $\Delta L = 2$ washout processes can be safely neglected in our analysis. These scattering processes, namely $LH \leftrightarrow \bar{L}H^{\dagger}$ (s-channel) and $LL \leftrightarrow HH$ (t- and u-channel), are mediated by the RHNs and have cross sections scaling as $\sigma \propto (Y_\nu Y_\nu^{\dagger})_{11}^{2}$, i.e.\ one power of the Yukawa coupling higher than the (inverse-)decay contribution, which scales as $(Y_\nu Y_\nu^{\dagger})_{11}$. This washout is therefore suppressed relative to the decay terms and becomes negligible for small couplings. Since our analysis has $(Y_\nu Y_\nu^{\dagger})_{11} \ll 1$, it can be safely neglected, becoming relevant only for Yukawa couplings of order unity.

However, in a more general case where a shift of regimes of thermal leptogenesis still leads to a flavored one, the quantum transport approaches (such as the closed-time path or density-matrix formalisms) is more appropriate \cite{DeSimone:2007gkc}. In this case, the CP asymmetry $\varepsilon_{1_{\alpha}}$ (for unflavored case, $\varepsilon_{1} = \sum_{\alpha} \varepsilon_{1_{\alpha}}$) is obtained from the decay of $N_1$ to a specific flavor $\ell_{\alpha}$ and estimated using standard expression~\cite{Covi:1996wh,Nardi:2006fx,DeSimone:2007gkc,Dev:2017trv}. The final baryon asymmetry $Y_{B}$ is then obtained by:
$Y_{B}= \frac{28}{79}\sum_\alpha n_{\Delta_\alpha}/s.$

\begin{figure}[h]
	\includegraphics[width=0.95\linewidth]{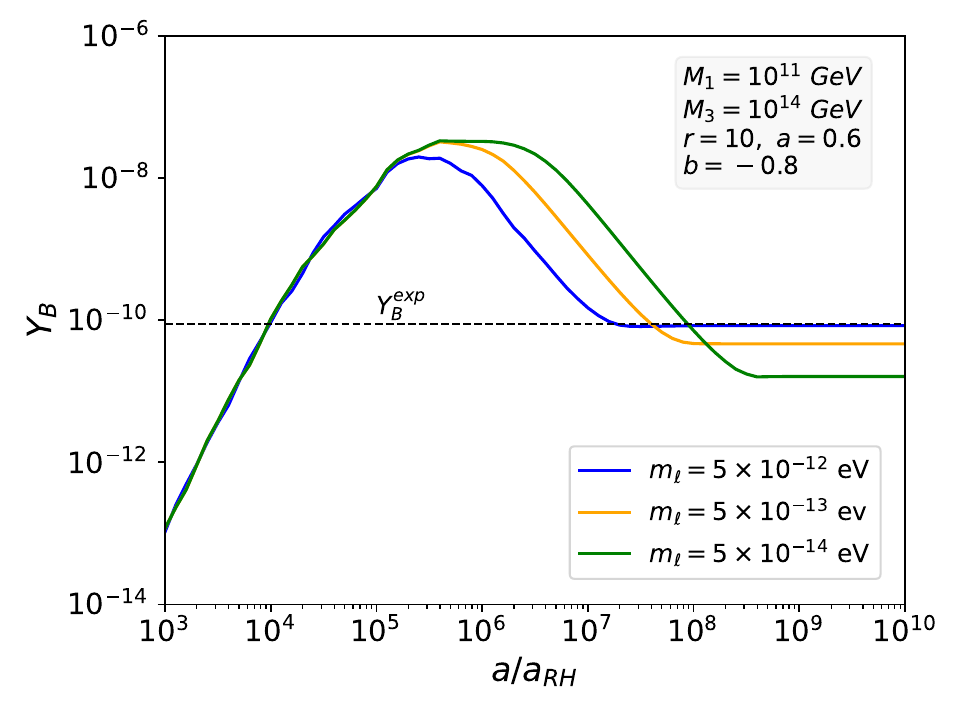}
	\caption{ Evolution of the baryon asymmetry with the expansion of the universe for a fixed value of parameters: $M_3=10^{14}~\rm{GeV},~M_1=10^{11}~\rm{GeV},~r=10,~a=0.6,~b=-0.8$ for three different values of $m_{\ell}$.}\label{asy}
\end{figure}

In Figure~\ref{asy}, we present the solution to the Eq. \eqref{BE_asy} solved simultaneously with the BEs of Eq. \eqref{BE} for a benchmark parameter set: $M_1=10^{11}$ GeV, $r=10$ ($M_2=rM_1$), $M_3=10^{14}$ GeV and $a=0.6,~ b=-0.8$. The values of $a$ and $b$ are chosen such that the observed baryon asymmetry $Y_B^{\rm exp}=8.75\times 10^{-11}$ \cite{Planck:2018vyg} is satisfied. Results are shown for three values of $m_\ell$, with $m_\ell=5\times 10^{-12}$ eV matching the observed value. The evolution proceeds in two distinct phases: first, the asymmetry from $N_1$ decay builds up until $N_1$ becomes Boltzmann-suppressed, marking the peak; second, entropy injection from the late-decaying $N_3$ dilutes the asymmetry. Additionally, the dependence on $m_\ell$ is also evident: smaller values extend the $N_3$-dominated era, leading to greater dilution of the asymmetry through enhanced entropy production from $N_3$ decay. Note that, the choice presented in Figure~\ref{asy} serves only as an illustrative example of how the modified expansion history delays the equilibration of the charged-lepton Yukawa interactions (shown in Figure~\ref{gammabyH}), enabling a comparatively lighter RHN ($M_1=10^{11}$ GeV) to produce the lepton asymmetry while preserving the quantum coherence among the different lepton-doublet states (unflavoured leptogenesis).

While we do not explicitly address experimental signatures in this work, the proposed scenario can, in principle, be tested through several channels. A key feature is that, within the conventional minimal type-I seesaw and without introducing any new field, sector, or symmetry, the heaviest RHN is automatically a long-lived state capable of driving an early MD epoch. The same Yukawa coupling of the decoupled RHN simultaneously fixes (i) its lifetime, (ii) the duration and entropy production of the MD epoch, and (iii) the lightest active-neutrino mass. The MD epoch is therefore not an independent input, but is tied directly to low-energy neutrino observables. As a concrete consequence, our scenario predicts a lightest neutrino mass $m_\ell \sim \mathcal{O}(10^{-12}-10^{-14})$ eV, rendering it falsifiable should ongoing or future experiments such as KATRIN \cite{Aker:2021exx} and the Project 8 collaboration \cite{Esfahani:2017dmu} succeed in probing this regime. Crucially, this correlation is absent in generic early-MD constructions, where the long-lived field and its decay properties are free parameters disconnected from any other observable sector. Additionally, studies such as \cite{Dror:2019syi, Chianese:2024gee, Datta:2024tne} suggest that it is indeed possible to probe leptogenesis through GW. In particular, such GWs can be sourced either by bremsstrahlung radiation from decaying heavy particles \cite{Khlebnikov:1997di, Kanemura:2023pnv, Datta:2024tne,  Konar:2025iuk} or by collapsing topological defects arising from spontaneous symmetry breaking \cite{Vilenkin:1981zs,Vachaspati:1984gt,Hiramatsu:2010yz,Roshan:2024qnv}. Detection of such a GW signal would unambiguously demonstrate that massive particles were produced and underwent decay in the early universe. On the other hand, the presence of a very heavy RHN in the early universe might also induce an era of early matter domination, which inevitably modifies all kinds of mentioned GW spectrums, including the same from the thermal plasma, known as the Cosmic Gravitational Microwave Background (CGMB) \cite{Ghiglieri:2015nfa, Ghiglieri:2020mhm, Ringwald:2020ist, Roshan:2024qnv, Murayama:2025thw}. Moreover, this would provide evidence that heavy particles once dominated the universe (see \cite{Murayama:2025thw} for details). An essential complementary GW signature may arise from cosmic string networks produced during the spontaneous breaking of an abelian gauge symmetry, which is responsible for the dynamical generation of RHN mass. The presence of early matter-dominated eras will lead to spectral breaks in the otherwise scale-invariant GW spectrum formed by cosmic strings \cite{Borah:2022byb, Borah:2022vsu, Borah:2023iqo}. The synergy between low-frequency GWs from cosmic strings and high-frequency signals at GHz and above will not only illuminate the cosmological history of leptogenesis but also probe a potential early MD era and the associated shift in charged lepton Yukawa equilibrium temperatures. 

To summarize, we have established that in the conventional minimal type-I seesaw mechanism, the heaviest decoupled RHN can induce an epoch of early MD, profoundly altering the universe's expansion history during periods when they dominate the cosmic energy budget. This modified expansion history has a significant impact on the equilibration of charged lepton Yukawa interactions. The implications of this phenomenon extend beyond the specific case studied here, encompassing a wider range of SM interactions and motivating further theoretical investigations. As a concrete application, we examine the consequences for leptogenesis arising from out-of-equilibrium production and decay of the lightest RHN. Additionally, we discuss various experimental avenues for testing this scenario, including prospects in neutrino physics and GW astronomy. While we restricted our discussion to the modification of charged lepton Yukawa interactions and its impact on leptogenesis, our results show that the neglected RHN of minimal type-I seesaw can play an important role in the early universe and affect other key processes, such as the production of dark matter, the origin of the matter-antimatter asymmetry, and the generation of primordial GWs.
\section*{Acknowledgment}			
	RR and SS would like to thank Partha Konar for the valuable discussions. RR acknowledges financial support from the STFC Consolidated Grant ST/T001011/1 and Short-Term Scientific Mission Grant provided by COST Action CA21136. RR also acknowledges the hospitality provided by Departamento de Fisica Fundamental and IUFFyM, Universidad de Salamanca where a part of this work was done. SS is supported by NPDF grant PDF/2023/002076 from the Science and Engineering Research Board (SERB), Government of India.
			
\bibliography{ref,ref2}
		
\end{document}